\documentclass[prl,twocolumn,a4paper,superscriptaddress,floatfix]{revtex4}
\usepackage{graphicx}
\begin{document}
\title{A unified model for vortex-string network evolution}

\author{C.J.A.P. Martins}
\email[Electronic address: ]{C.J.A.P.Martins@damtp.cam.ac.uk}
\affiliation{Centro de Astrof\'{\i}sica da Universidade do Porto, R. das
Estrelas s/n, 4150-762 Porto, Portugal}
\affiliation{Department of Applied Mathematics and Theoretical Physics,
Centre for Mathematical Sciences,\\ University of Cambridge,
Wilberforce Road, Cambridge CB3 0WA, United Kingdom}
\affiliation{Institut d'Astrophysique de Paris, 98 bis Boulevard Arago,
75014 Paris, France}
\author{J.N. Moore}
\email[Electronic address: ]{J.N.Moore@damtp.cam.ac.uk}
\affiliation{Department of Applied Mathematics and Theoretical Physics,
Centre for Mathematical Sciences,\\ University of Cambridge,
Wilberforce Road, Cambridge CB3 0WA, United Kingdom}
\author{E.P.S. Shellard}
\email[Electronic address: ]{E.P.S.Shellard@damtp.cam.ac.uk}
\affiliation{Department of Applied Mathematics and Theoretical Physics,
Centre for Mathematical Sciences,\\ University of Cambridge,
Wilberforce Road, Cambridge CB3 0WA, United Kingdom}

\date{21 October 2003}

\begin{abstract}
We describe and numerically test the velocity-dependent one-scale (VOS)
string evolution model, a simple analytic approach describing a string
network with the averaged correlation length and velocity.   
We show that it accurately reproduces the large-scale behaviour (in particular
the scaling laws) of numerical simulations of both Goto-Nambu and field theory
string networks.  We explicitly demonstrate the relation between the high-energy
physics approach and the damped and non-relativistic limits which are relevant
for condensed matter physics.  We also reproduce experimental results in this
context and show that the vortex-string density is significantly reduced by loop
production, an effect not included in the usual `coarse-grained' approach. 
\end{abstract}
\preprint{DAMTP-2000-32}
\maketitle

%%%%%%%%%%%%%%%%%%%%%%%%%%%%%%%%%%%%%%%%%%%%%%%%%%%%%%%%%%%%%%%%%%%%%%%%%%
\section{\label{sint}Introduction}

Vortex-lines or topological strings can appear in a wide range of physical 
contexts, ranging from cosmic strings in the early universe to vortex-lines
in superfluid helium (for reviews see ref.\cite{vsh,cond1,cond2}).
Gaining a quantitative understanding of their important effects
represents a significant challenge because of their nonlinear nature and 
interactions and because of the complexity of evolving string networks.  
Considerable
reliance, therefore, has been placed on numerical simulations but 
unfortunately these turn out to be technically difficult and very 
computationally costly. This provides strong motivation for alternative
analytic approaches, essentially abandoning the detailed `statistical 
physics' of the string network to concentrate on its `thermodynamics'.
Here we present one such model for string network evolution, the velocity-dependent one-scale (VOS) model \cite{ms2,model}, and demonstrate 
its quantitative 
success by direct comparison with numerical simulations.  We are able to 
describe the scaling laws and large-scale properties of string networks 
in both cosmological and condensed matter settings.

The first assumption in this analysis is to `localise' the string so that we 
can treat it as a one-dimensional line-like object.  This is clearly a good
assumption for gauged strings, such as magnetic flux lines, but may seem
more questionable for strings possessing long-range interactions, such as 
global strings or superfluid vortex lines.  However, as we shall see, 
we will be able to establish good agreement between the VOS model and 
simulations in both `local' and `global' cases.  
The second step is to 
average the microscopic string equations of motion to derive the key 
evolution equations for the average string velocity $v$ and correlation 
length $L$.  This is a generalization of Kibble's original `one-scale'
model \cite{kib}, and has been described elsewhere  \cite{model}. 

We make a detailed comparison between the VOS model and numerical simulations -- currently the world's largest and
highest resolution -- using both direct field 
theory simulations of magnetic flux-lines, as well as simulations of Nambu
strings, treating them 
as localised line-like objects.  We are able to demonstrate good agreement 
between all three approaches, thus underpinning the important assumptions
required for the VOS model.  Fixing a single parameter, we are able to 
provide a good description of cosmic strings throughout the history of the universe, from the friction-dominated regime after their 
formation, through the radiation-matter transition and into the accelerating epoch today. Significantly, we believe the model also describes the same average
features of an evolving vortex-line tangle in a condensed matter context, 
reproducing the expected and experimentally observed scaling law.  The 
present VOS model in this context averages over both the background 
fluid friction and the Magnus force, but we believe it can be adapted 
further to incorporate other physical effects. 

%%%%%%%%%%%%%%%%%%%%%%%%%%%%%%%%%%%%%%%%%%%%%%%%%%%%%%%%%%%%%%%%%%%%%%%%%%

\section{\label{smod}The VOS string network model}

It is interesting to contrast the condensed matter and high-energy physics
approaches to vortex string dynamics. In high-energy physics one uses,
whenever possible, a one-dimensional view (obtained by integrating over
radial modes of the vortex solution), described by the Nambu action.
In this approximation the string equations of motion are \cite{ms2,condmat}
\begin{eqnarray}
&& \ddot {\bf x}
+ \left(1-{\dot {\bf x}}^2\right)\,\frac{\dot {\bf x}}{\ell_d} =
\frac{1}{\epsilon} \,
\left(\frac{{\bf x}^\prime}{\epsilon}\right)^\prime+\zeta{\dot {\bf x}}
\wedge\frac{{\bf x}^\prime}{\epsilon}
\label{strings}\\
&&\dot \epsilon + \frac{\dot {\bf x}^2}{\ell_d}\,\,\epsilon = 0\,
\label{stringt}
\end{eqnarray}
where ${\bf x}$ is the string position, $\epsilon$ is a measure of the energy
along the string, and $\ell_d=2 {\cal H}+\mu/\beta T^3$ is a
damping length scale;
time derivatives are with respect to conformal time
(for condensed matter purposes this is identical to physical time).
All terms can be rigorously derived from the Nambu action, except the last 
one in (\ref{strings}) which comes from the Kalb-Ramond action and
describes the Magnus force arising when a global string
moves through a relativistic superfluid background \cite{DavisShell} whose density is parametrized by $\zeta$ \cite{condmat,sch2}.  

Now let us consider two limits of the equations of motion that are relevant
in condensed matter systems.
Firstly, if the damping term dominates, we find after some algebra (note
that the Magnus force is not included, since it's not dissipative)
\begin{equation}
\frac{\dot {\bf x}}{\ell_d} =-\frac{1}{{\bf x}^{\prime4}}\left[
{\bf x}^\prime\wedge({\bf x}^\prime\wedge{\bf x}^{\prime\prime})\right]\,,
\label{limit1}
\end{equation}
where the right-hand side is the friction force term, e.g., on a superfluid 
vortex \cite{sch1,sch2,DavisShell,condmat}. Secondly, let us consider the
non-relativistic limit but without damping. In this case we find
\begin{equation}
\dot {\bf x}=\frac{1}{\zeta{\bf x}^{\prime2}}\frac{{\bf x}^\prime\wedge{\bf x}^{\prime\prime}}{\epsilon}\,,
\label{limit2}
\end{equation}
which is the equation describing the frictionless motion of a vortex
filament in an unbounded fluid \cite{sch1,sch2}.

By combining these two terms we can therefore reproduce the equation of motion
obtained by Schwarz \cite{sch1,sch2}, which used an effective and
more phenomenological 1D approach (based on a coarse-grained order
parameter) to obtain the terms one by one. Note however that the
Schwarz equation has further (subdominant) terms, coming from non-local
and boundary contributions (which we have neglected altogether).

Now, it is well known experimentally \cite{cond1,cond2} that vortices in
such systems evolve following a scaling law $L\propto t^{1/2}$, and the same 
is true for example for defects in liquid crystals \cite{chuang}. In fact
this is expected to hold for any phase ordering with a non-conserved
order parameter (and in this context it is sometimes referred to as the
Lifshitz-Allen-Cahn equation). We will see below that this scaling law can 
be easily derived from an averaged
version of the evolution equations (\ref{strings}-\ref{stringt})---a process
that might be compared to coarse-graining---in addition to also being
relevant in high-energy contexts.

The VOS model has been described in
detail elsewhere \cite{ms2,model},
so here we limit ourselves to a brief summary. Any string
network divides fairly neatly into two distinct populations, long 
(or `infinite') strings and small closed loops.  
A phenomenological term must then be included to account for the loss of
energy from long strings by the production of loops, which are much smaller
than $L$---this is the `loop chopping
efficiency' parameter $\tilde c$. A further phenomenological term (characterized
by a strength $\Sigma$ and a characteristic length scale $L_d$) is also
included to account for radiation back-reaction effects. Note that all these
parameters are constants---refer 
to \cite{ms2,model,msm} for details. This term is negligible for 
GUT-scale Goto-Nambu string networks, but can be relevant in other cases.

By suitably averaging Eqns. (\ref{strings}--\ref{stringt}) one can 
obtain the following evolution equations
\begin{equation}
2\frac{dL}{dt}=2HL+{\tilde c}v+\frac{L}{\ell_d}v^2+8\Sigma
v^6\exp\left(-\frac{L}{L_d}\right)\, , \label{evl}
\end{equation}
\begin{equation}
\frac{dv}{dt}=\left(1-{v^2}\right)
\left(\frac{k(v)}{L}-\frac{v}{\ell_d}\right)\, ;
\label{evv}
\end{equation}
here $H$ is the Hubble parameter (relevant for cosmology) and $k$ is a
velocity-dependent phenomenological (but otherwise universal) parameter, called the `momentum parameter', given by \cite{model}
\begin{equation}
k(v)=\frac{2\sqrt{2}}{\pi}(1-v^2)(1+2\sqrt{2}v^3)\frac{1-8v^6}{1+8v^6}\label{defnk}\,;
\end{equation}
its non-relativistic limit is $ k_{\rm nr}(v)\sim2\sqrt{2}/\pi$.

We can now take the condensed matter (non-relativistic) limit of the VOS model.
All we need to do is set $H=0$ and the damping length to a constant. One finds
a stable attractor solution
\begin{equation}
L=\sqrt{1+{\tilde c}}\left(\ell_d t\right)^{1/2}\, . \qquad
v=\frac{\ell_d}{L}\label{scal}\,.
\end{equation}

We have thus shown that our Goto-Nambu based microscopic equations of
motion and our averaged version of them successfully reproduce known condensed
matter results, respectively the Schwarz equation and the $L\propto t^{1/2}$
scaling law. Note that our solution demonstrates the importance of loop
production---although this is not usually included in theoretical or numerical
analyses in the condensed matter context, it has been observed in 
experiments \cite{chuang}. We will now further test our averaged model
in the context of field theory simulations and cosmology.

\begin{figure}
\includegraphics[width=3in]{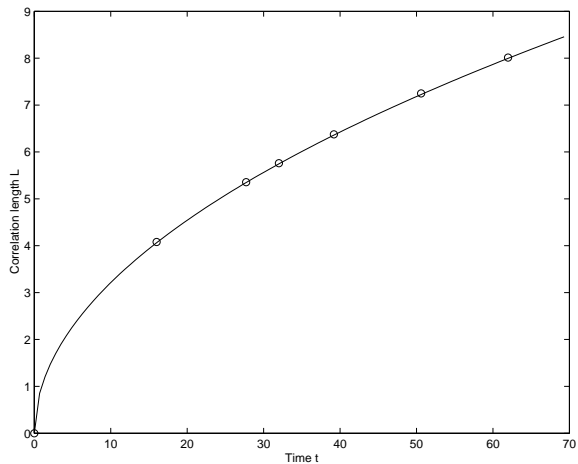}
\caption{\label{figfric}Friction-dominated evolution of a vortex-string network.
Measured values of the average correlation length during
diffusive evolution of an Abelian-Higgs string simulation are plotted against
the best-fit scaling law $L\propto t^{1/2}$ (\protect\ref{scal}).}
\end{figure}

\section{\label{sfield}Field theory simulations}

In a previous paper \cite{msm} we have described the evolution of 
string networks in full 3D field theory simulations of 
the Abelian-Higgs model.  This is a relativistic analogue of the
Ginzburg-Landau theory of superconductivity.  Before  reviewing the 
comparisons between these simulations and the VOS model, it is 
worthwhile noting a byproduct of this work which relates
to vortex tangles in condensed matter.  In order to create quiescent initial
conditions for string evolution in these simulations, instead of the 
relativistic equations, we began by solving the corresponding diffusive equations (refer to \cite{msm} for details).
This was essentially equivalent to the 
non-relativistic evolution of magnetic flux-lines in a friction-dominated 
regime.  Measurements of the string correlation length for the evolving network 
revealed a clear $L \propto t^{1/2}$ scaling, as
illustrated in Fig. \ref{figfric}.  As well as confirming the VOS model
prediction in this case (\ref{scal}), this has already been observed experimentally \cite{chuang}.

\begin{figure}
{\hbox{\includegraphics[width=1.75in]{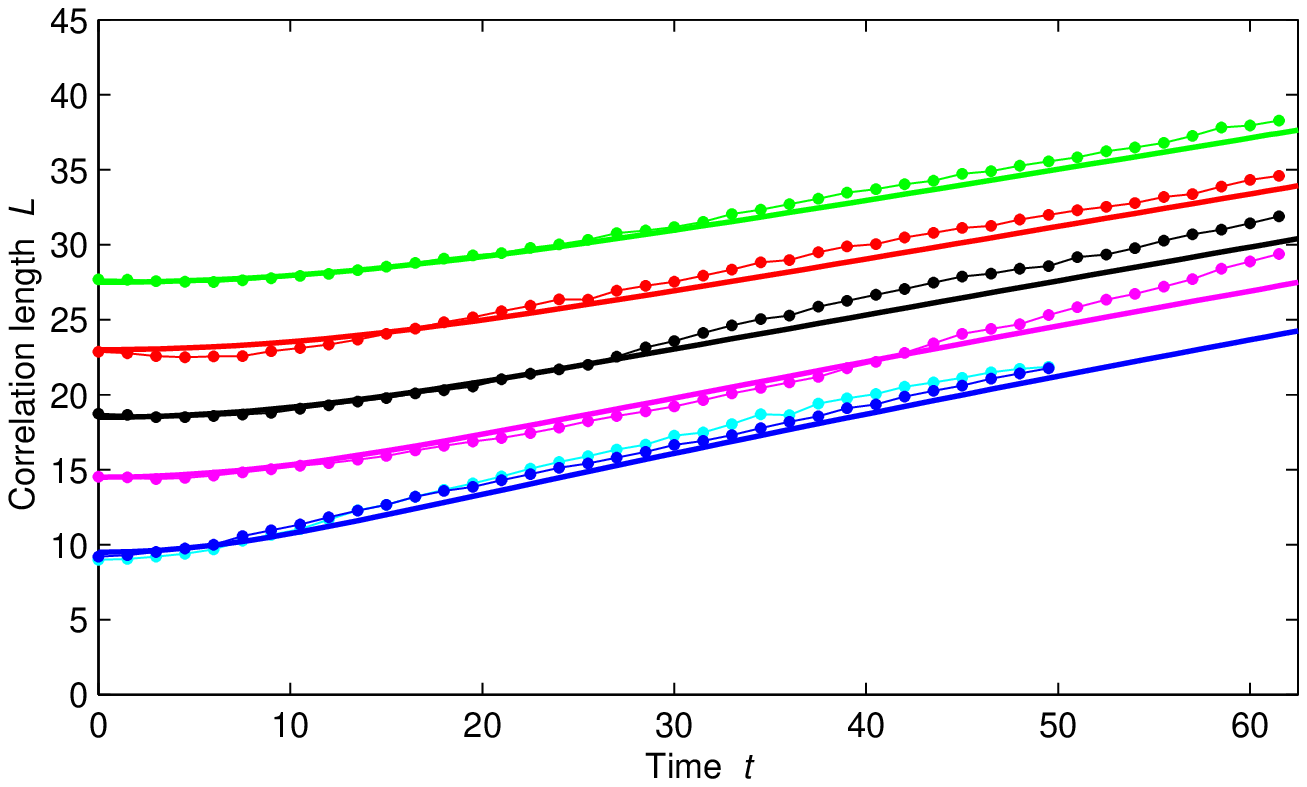}
\includegraphics[width=1.75in]{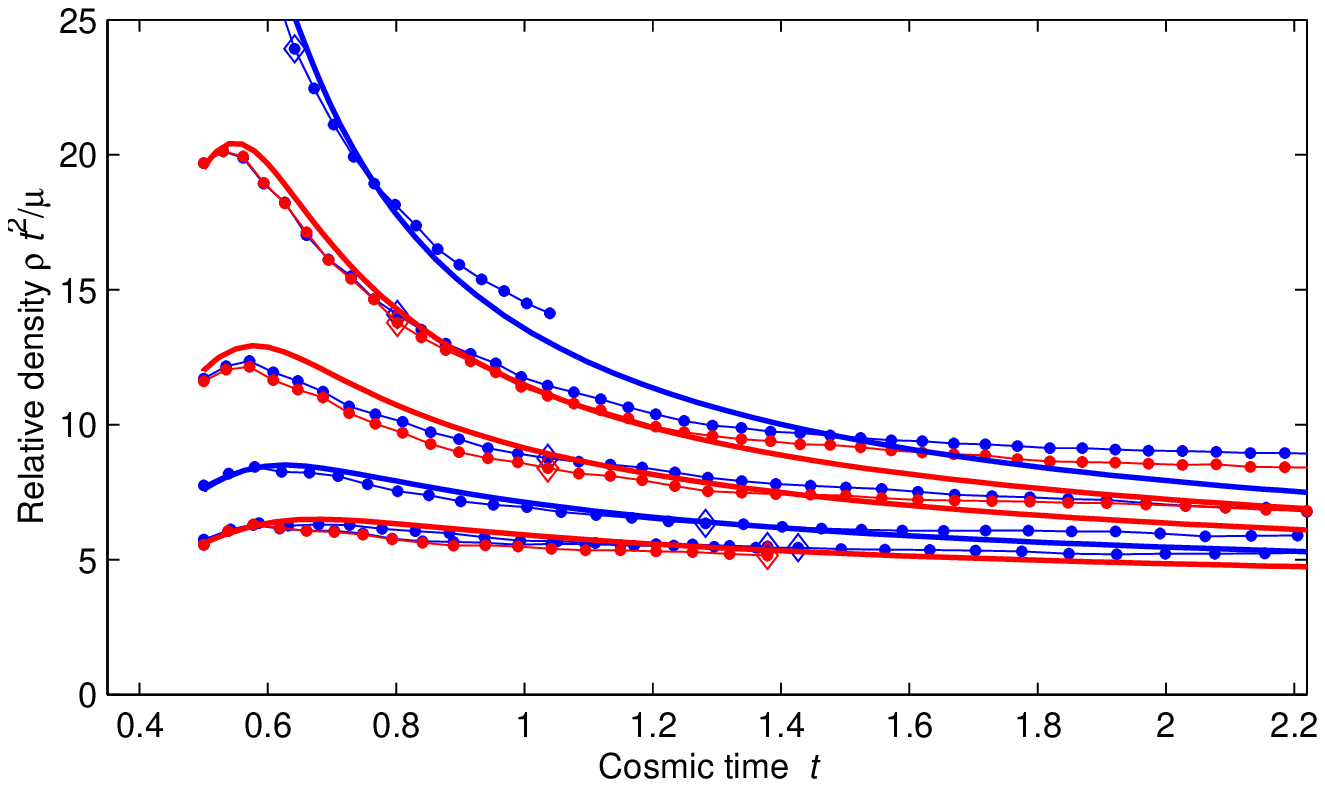}}}
{\hbox{\includegraphics[width=1.75in]{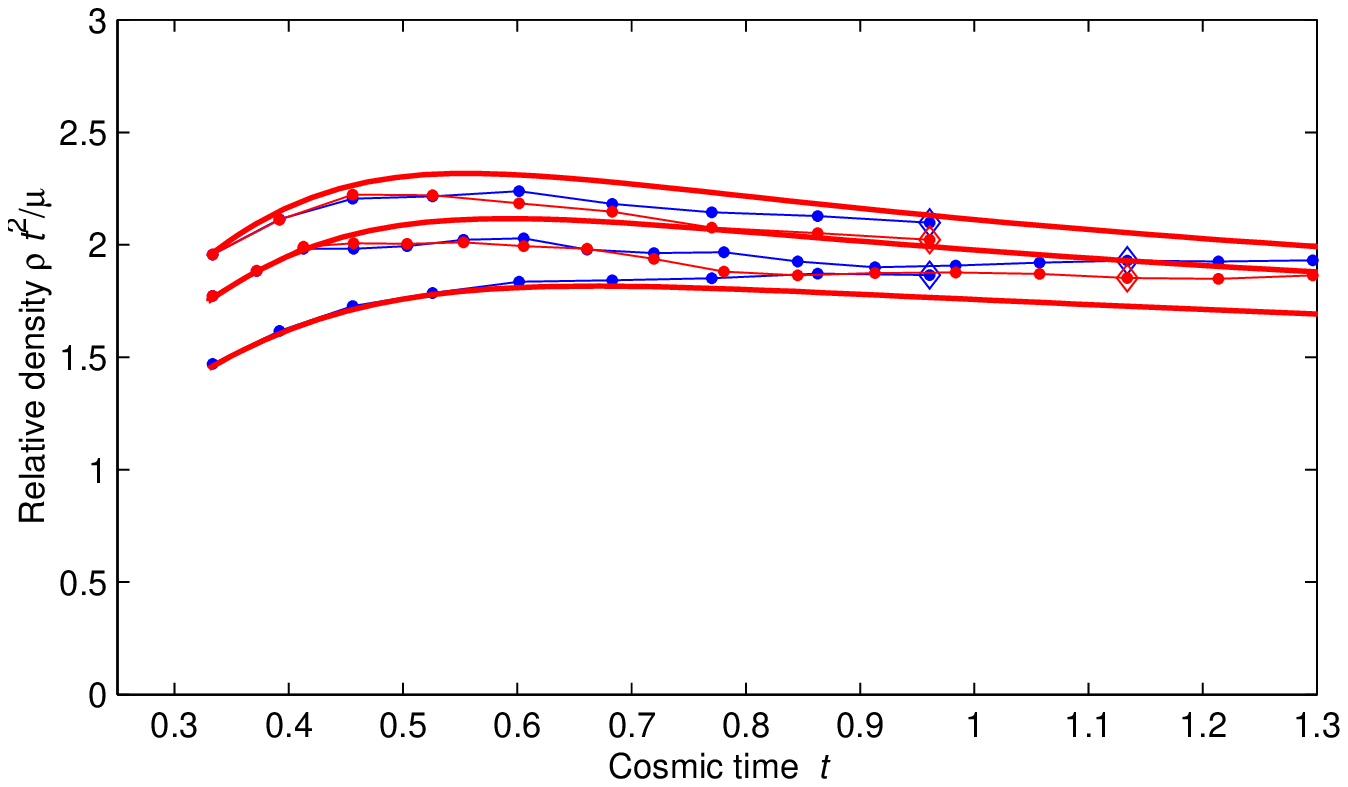}
\includegraphics[width=1.75in]{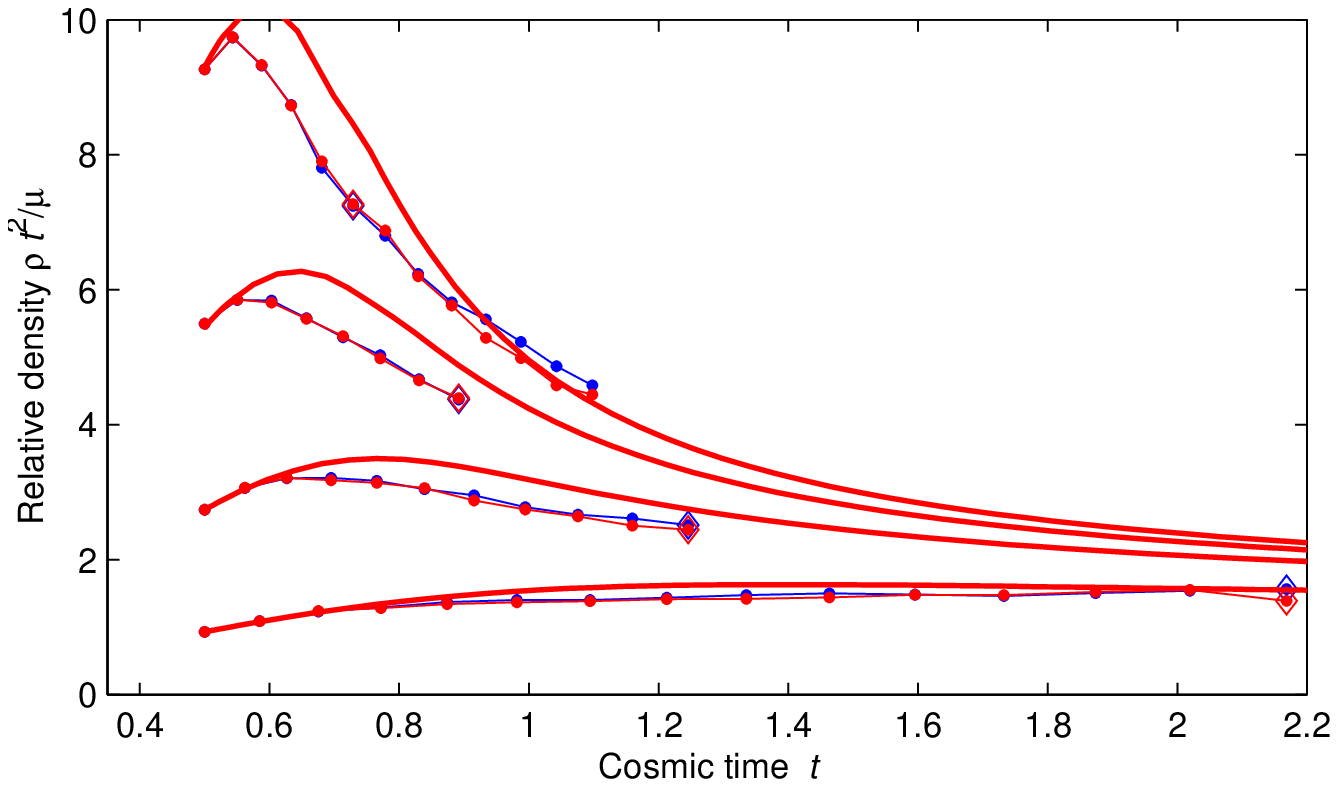}}}
\caption{\label{field} Fitting the VOS model to flat (top left), radiation
(top right), matter (bottom left) and global (bottom right)
field theory simulations, with ${\tilde c}=0.57$,
$\Sigma=0.5$ (except that for global it's 1.1) and $L_d=4\pi$.}
\end{figure}

The relativistic evolution of the string networks also clearly revealed the
predicted scaling laws and, remarkably, the correlation length and velocities 
for all the simulations had a good asymptotic fit from the VOS model using
the single parameter $\tilde c \approx 0.57$.  This fit was universal
regardless 
of whether the simulations were in flat space or in an expanding universe, or 
whether matter or radiation eras, as illustrated in fig.~\ref{field}.  There is no degeneracy here between the chopping efficiency $\tilde c$ which determines the asymptotics and the massive radiation parameters $\Sigma = 0.5$ and $L_d = 4\pi$ which only affect the initial conditions.  However, for global strings with massless radiation $L_d\rightarrow \infty$
there is such a degeneracy between $\tilde c$ and $\Sigma$ because they act in 
the same manner asymptotically.  However, assuming the same loop chopping efficiency $\tilde c = 0.57$ for local and global cases, requires a much higher damping
coefficient $\Sigma = 1.1$ for the latter as expected for massless radiation \cite{msm}. (These results and fits are also in reasonable agreement with
other recent simulations of global strings in ref.~\cite{yamaguchi}.)  
This excellent correspondence for both 
local and global strings appears to 
establish the validity of the two key ('localization' and 'thermodynamic') 
assumptions underlying the VOS model.

\section{\label{hires}High-resolution gauge string simulations}

\begin{figure}
\includegraphics[width=3in]{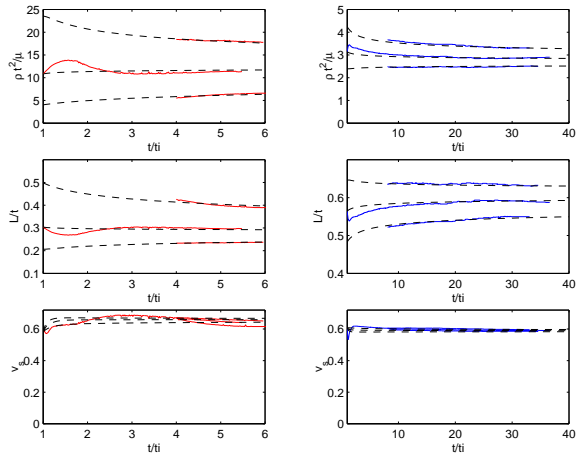}
\caption{\label{radmat} The approach to scaling of radiation (left) and
matter (right) era ultra-high resolution
simulations. Solid lines are data from
three different runs (for each epoch). All dashed lines are solutions of
the analytic model, with the relevant initial conditions, for ${\tilde c}=0.23$.}
\end{figure}

We now compare our model with ultra-high resolution numerical simulations,
the details of which will be reported elsewhere. Specifically,
we discuss both the approach to the linear scaling regime in the radiation and
matter epochs, and the transition between the two epochs. Note that in this
case both friction and radiative back-reaction are negligible.
We first compare the model with three different ultra-high resolution runs
(75 points per correlation length, fixed physical resolution) in the
radiation and matter epochs, see
Fig. \ref{radmat}. For all 6 runs, the analytic model fit displayed corresponds
to a loop chopping efficiency parameter ${\tilde c}=0.23$.
We find that this value also approximately reproduces the earlier results of
Bennett and Bouchet \cite{bb} and Allen and Shellard \cite{as}. Residual
differences can be attributed to the much higher resolution of our runs.
Earlier work \cite{at} reported different scaling
values, but numerical codes were substantially improved subsequently.
Thus fixing this parameter via the radiation era, our model correctly predicts
the matter era scaling large-scale properties without any further tampering
with parameters.We estimate the loop chopping efficiency to have the value
${\tilde c_{ren}}=0.23\pm0.04$.
We emphasize that we expect this to be a `universal' parameter, independent of
the cosmological scenario in which the string network is evolving. However,
if one performs analogous simulations in flat (Minkowski) spacetime, one does
find a different value, ${\tilde c_{ren}}=0.57\pm0.04$, which coincides with
the value we found above for field theory simulations. This is because the
amount of small scale structure present in Goto-Nambu expanding runs is
much larger than that in field theory and/or Minkowski space runs, and this
has an influence on the large-scale features of the network described by
the model. Hence the two values can be regarded as the `renormalized'
and `bare' chopping efficiencies.  This is discussed at greater length
in \cite{msm}.

\begin{figure}
\includegraphics[width=3in]{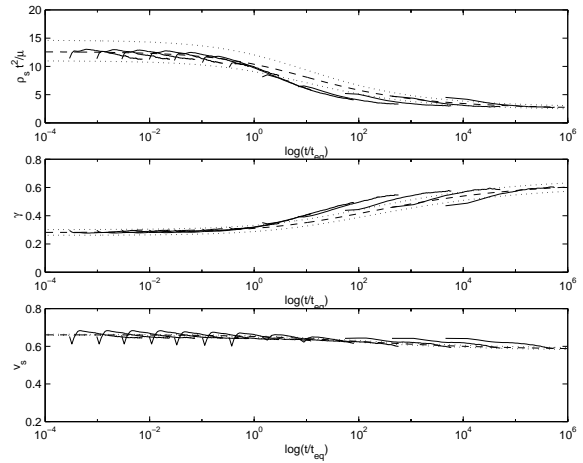}
\caption{\label{trans} The evolution of a GUT-scale string network in the
transition from the radiation to the matter-dominated epochs. Solid lines are
data from a set of 12 different high-resolution runs.
The dashed curve is the solution of the VOS model with ${\tilde c}=0.23$,
while the dotted curves correspond to  ${\tilde c}=0.21$ and ${\tilde c}=0.25$.}
\end{figure}

Now, given that we can reproduce the radiation and matter era scaling
values, can we also reproduce the transition between them?
Note that this point is not
straightforward: nothing guarantees us, {\it a priori}, that the model
will produce the correct timescale for the transition even if both its
asymptotic limits are correct.
Here it is not possible to do a single numerical simulation
spanning all the relevant time interval, so we have performed a
sequence of twelve different high-resolution runs
(16 points per correlation length, fixed physical resolution) which
together span the required range, see Fig. \ref{trans}.
Again we can see that the analytic
model, with the previously determined parameter ${\tilde c}=0.23$ can
actually do a good job in predicting the overall timescale of the
transition between the regimes. 
We also emphasise that a GUT-scale network at the present time has not quite
yet reached the asymptotic matter-era scaling regime. This slow relaxation
process has been pointed out before \cite{ms2}, and is clearly of vital
importance for an adequate analysis of cosmic string structure formation
scenarios.

%%%%%%%%%%%%%%%%%%%%%%%%%%%%%%%%%%%%%%%%%%%%%%%%%%%%%%%%%%%%%%%%%%%%%%%%%%

\section{\label{sdsc}Discussion and conclusions}

We have described a simple analytic model for vortex-string evolution and
shown how it successfully reproduces the large-scale features of numerical simulations of both Goto-Nambu and field theory
string networks, as well as of experiments in condensed matter physics.
This quantitative correspondence provides strong evidence in support of
the main assumptions on which the VOS model is based,
notably string `locality' and 
`thermodynamic' averaging.  Our results confirm that the dominant mechanisms
affecting string network evolution are loop production and damping, whether
from friction or radiation depending on the context. In condensed matter
systems we find that the two are of comparable magnitude, despite loop
production being neglected in the usual treatments. For global strings,
loop production is dominant but radiative damping can significantly affect 
the network density.  The outstanding issue raised by comparisons of 
gauged string networks in Nambu and field theory simulations remains the modelling of small-scale features.  A more detailed analysis of the simulations provides some valuable insights into this problem, but 
this will be the subject of a forthcoming publication.  

\section{acknowledgments}
We would like to thank Bruce Allen, Pedro Avelino, Brandon Carter,
Levon Pogosian, Tanmay Vachaspati and Proty
Wu for useful conversations. Our Nambu simulations used the
Allen-Shellard string network code \cite{as}.
C.M. is funded by FCT
(Portugal), under grant no. FMRH/BPD/1600/2000.
This work was done in the context of the ESF COSLAB network, and
performed on COSMOS, the Origin3800 and Altix3700 owned by the UK
Computational Cosmology Consortium, supported by SGI, HEFCE and PPARC.

%%%%%%%%%%%%%%%%%%%%%%%%%%%%%%%%%%%%%%%%%%%%%%%%%%%%%%%%%%%

\bibliography{testing4}

\end{document}